\documentclass[a4paper]{jpconf}
\usepackage{graphicx}
\usepackage{paralist}

\begin{document}
\title{On the nature of the  $\delta$ Scuti
star HD 115520}

\author{J. H. Pe\~na$^1$, L. Fox Machado$^2$, B. Cervantes-Sodi$^1$, R. Pe\~na$^3$, G. Mu\~noz$^4$, B. Vargas$^4$, J. P. Sareyan$^5$, M.
Alvarez$^2$, M. Cano$^1$, M. A. Sorcia$^1$}

\address{
$^1$ Instituto de Astronom\'{\i}a, Universidad Nacional Aut\'onoma
de M\'exico, M\'exico D.F., Apdo. Postal 70-264,
M\'exico\\
$^2$ Observatorio Astron\'omico Nacional,
Instituto de Astronom\'{\i}a,
Universidad Nacional Aut\'onoma de M\'exico, Ensenada B.C., Apdo.
Postal 877,
M\'exico\\
$^3$Facultad de Ciencias, Universidad Nacional Aut\'onoma de M\'exico,
M\'exico D.F., M\'exico\\
$^4$ Escuela Superior de Ingenier\'{\i}a, Mec\'anica y El\'ectrica,
Instituto Polit\'ecnico Nacional, Av. IPN s/n, 07738 M\'exico, D.F.,
M\'exico\\
$^5$Observatoire de la Cote d'Azur, France}

\ead{jhpena@astroscu.unam.mx}

\begin{abstract}
Observing Delta Scuti stars is most important as their
multi-frequency spectrum of radial pulsations provide strong
constraints on the physics of the stars interior; so any new
detection and observation of these stars is a valuable contribution
to asteroseismology. While performing uvby-beta photoelectric
photometry of some RR Lyrae stars acquired in 2005 at the
Observatorio Astron\'omico Nacional, M\'exico, we also observed
several standard stars, HD115520 among them. After the reduction
this star showed indications of variability. In view of this, a new
observing run was carried out in 2006 during which we were able to
demonstrate its variability and its nature as a Delta Scuti star.
New observations in 2007 permitted us to determine its periodic
content with more accuracy. This, along with the uvby-beta
photoelectric photometry allowed us to deduce its physical
characteristics and pulsational modes.
\end{abstract}

\section{Introduction}
In [1] confirmed  the membership of HD 115520 to the $\delta$ Scuti
class. It had been considered as a standard star in a 2005 observing
run. From the relatively large scatter shown, [2] considered it to
be variable candidate. With this in mind, new data were acquired in
two new nights in 2006 which established it as a $\delta$ Scuti
star. In the present paper we present new observations which were
performed in 2007 with the same instrumentation over a period of
four nights and which have served to determine its periodic content.
The frequencies found explain the behavior of both seasons separated
by more than one year.

\section{Observations}

These were taken at the Observatorio Astr\'onomico Nacional,
M\'exico using the 1.5 m telescope to which a spectrophotometer was
attached. The observing season was carried out on four consecutive
nights in March and April, 2007. The following observing routine was
employed: a multiple series of integrations was carried out,
consisting of five 10 s integrations of the star to which one 10 s
integration of the sky was subtracted. Two reference stars were also
observed C1: HD116879 and C2: HD114311. These were observed in the
following sequence to optimize the time coverage of the variable: V,
sky, C1, V, V, C2, V. A series of standard stars was also observed
at the beginning and at the end of each night to transform the data
into the standard system. The Str\"omgren system [3] is an
intermediate band width system that overcomes many of the
shortcomings of the $UBV$ system and provides astrophysically
important information. The color indices in the Str\"omgren system
are very useful quantities. Because both the $b$ and $y$ filters are
relatively free from blanketing, the index $(b-y$ is a good
indicator of color and effective temperature. A color index is
essentially the slope of the continuum. In the absence of
blanketing, the continuum slope would be roughly constant and
$(b-y)$ approximately equals $(v-b)$. Because $(v-b)$ is affected by
blanketing, the difference between these two indices indicates the
strength of blanketing. Hence a metal index, $m_{1}$, can be defined
as

\begin{equation}
m_{1} = (v-b) - (b-y).
\end{equation}

To determine how the continuum slope has been affected by the Balmer
discontinuity, the index $c_{1}$ is definde as

\begin{equation}
c_{1} = (u-v) - (v-b).
\end{equation}

This index measures the Balmer discontinuity, nearly free from the
affects of line blanketing.

The absolute photometric values of the 2007 campaign are provided in
an archive. The accuracy of the season is deduced from the
differences between the reduced and the previously reported values
of the standard stars. Due to the fact that the last night was of
lower quality, and hence, less accurate, the mean values of the
differences are calculated only from the standards of the first
three nights. They are: 0.015, 0.008, 0.007, 0.011 mag for V,
$(b-y)$, $m_1$, and $c_1$, respectively.

However, since the amplitude of the star is typical of a  $\delta$
Scuti star ($\sim$ 20 mmag, see Figure 4 and in Figure 2 of [1]), we
preferred to analyze the data for the periodic content through
differential photometry in the $y$ filter for which use was made of
the reference  stars C1 and C2 to increase the accuracy of the
photometry to thousands of magnitude. A magnitude value of the
reference stars was interpolated at the time of the variable and the
final values, to which the average value of each night was
subtracted.

\section{Frequency determination}
With the relatively few data points acquired in the 2006 season
(only two short nights) we were able to demonstrate the star's
variability and found evidence of at least two close frequencies
which might explain the resulting beating behavior of the light
curve. Since the new photometric data is constituted of four long
consecutive nights, we are now able to determine  the pulsational
frequencies with greater precision. Two numerical packages were
utilized: Period04 [4] and ISWF [5]. With Period04 the first run
examined gave a frequency of 17.8643 c/d with an amplitude of 0.0140
mag in the frequency interval between 0 and 30 c/d with a step rate
of 0.0150. On the other hand, the ISWF package yielded the following
frequencies (in c/d) listed in diminishing amplitudes (in
parentheses, in mmag) 17.850 (13.877); 14.7786 (10.334); 17.4527
(6.415); 13.5217 (4.236) and 18.1831 (3.973).

In the 2006 season we obtained 18.82 and 14.63 c/d. Given the
complex window function of observations on only two nights from only
one observatory, we might consider them the same. On the other hand,
when the whole set was utilized with a step rate of 0.00015,
Period04 yielded peaks at 17.8373 and 14.7537 c/d, (see Figure 1 and
Table 1). The rest of the frequencies might be disregarded because
they do not significantly improve the residuals. Their  peaks are
indistinguishable from each other due to the aliasing caused by the
window function. Therefore, we will consider as definitive only the
first two frequencies listed in Table 1. Figure 2 shows the light
curves of the six observed nights.

\begin{figure}[t]
\begin{center}
\includegraphics[width=10cm]{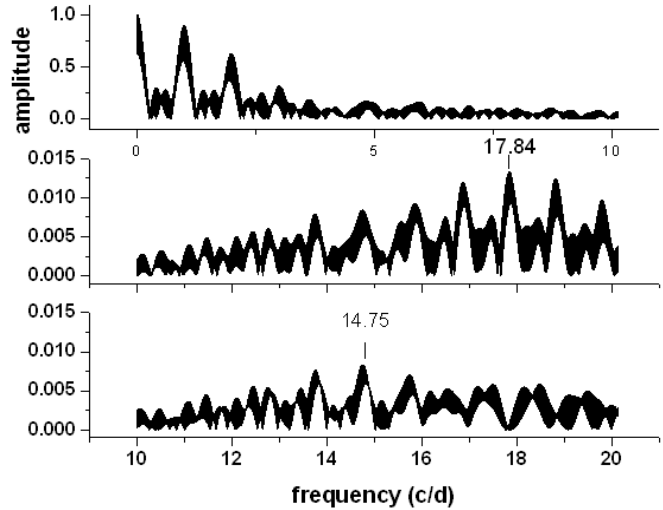}
\caption{Periodograms of all the observed nights. From top to
bottom, window, first frequency obtained at 17.8375c/d, periodogram
after prewhitening this with a resulting peak at 14.7537c/d, and
finally the prewhitened histogram of the two previously determined
frequencies with a peak at 16.5121c/d.}
\end{center}
\end{figure}

\begin{table*}[!t]\centering
  \setlength{\tabcolsep}{1.0\tabcolsep}
 \caption{ Frequencies, amplitudes and phases derived }
  \begin{tabular}{llllll}
\hline\hline
 & Frequency (c/d)    &    Amplitude (mag) & Phase    \\
\hline
F1 & 17.8375  &  0.0131 &  0.1028\\
F2 & 14.7537  &  0.0108 &  0.2612\\
F3 & 16.5121  &  0.0070 &  0.5646\\
\hline
\end{tabular}
\end{table*}

\begin{figure}[t]
\begin{center}
\includegraphics[width=10cm]{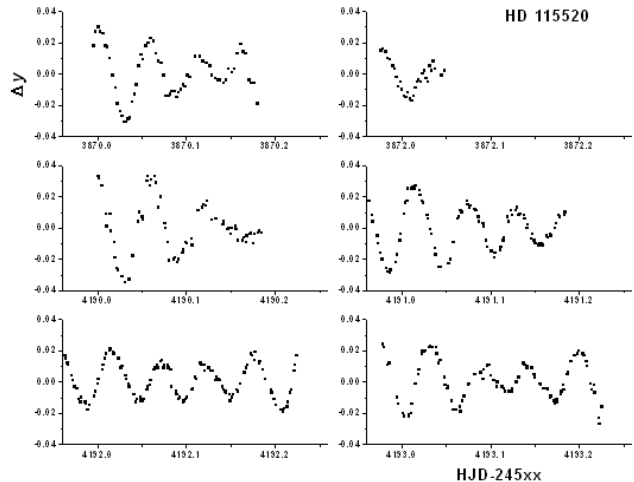}
\caption{Light curves of HD 115520 for the years 2006 and 2007
obtained at SPM Observatory. $y$ shows the magnitude variation
(dots) in magnitudes, X axis is time. }
\end{center}
\end{figure}

\section{Physical parameters}
As it has been already described in [1], we carried out a well-known
procedure to determine reddening as well as unreddened colors using
the photometric mean $uv$ values reported in Table 3. Table 4 lists
the reddening, the unreddened indexes, the absolute magnitude, and
the distance. Its position on the $[m_1]-[c_1]$ diagram established
it to be an A8V star. Its temperature and log of surface gravity can
be determined by locating HD 115520 in the $(b-y)_0$ vs. $c_0$ grids
of [6] (Figure 3); the values we determine are 7700 K and 4,
respectively. As was stated in Paper I, we compared our results with
those in a paper by [7] who found  an effective temperature $T_{\rm
eff}$ of 8199 (+449,-317), a log g 4.63 (+0.34,-0.23), an [Fe/H]
0.62 (+-0.13) and a stellar type membership to the main-sequence for
this star. Although [7] has evaluated physical parameters for this
star, and his numerical values coincide with ours, we feel that we
have more data to determine the physical characteristics.
Nevertheless, we have employed his reported metallicity of HD 115520
to discriminate between the models that explain the star's behavior.

\begin{table*}[!t]\centering
  \setlength{\tabcolsep}{1.0\tabcolsep}
 \caption{ Mean values of the $uv$ photometry of HD 115520 from the two seasons }
  \begin{tabular}{lllllllll}
\hline\hline
 & average &   sigma     &    N     \\
\hline
V (mag)      & $8.4305$ & $0.0178$ & $579$ \\
$(b-y)$ (mag) & $0.1334$ & $0.0070$ & $584$ \\
$m_1$   & $0.1701$ & $0.0051$ & $580$ \\
$c_1$   & $0.8068$ & $0.0139$ & $584$\\
$\beta$ & $2.8108$ & $0.0133$ & $67$ \\
\hline
\end{tabular}
\end{table*}

\section{The evolutionary status of HD 115520}

The determination of the evolutionary stage of a field star requires
precise estimates of its global parameters. In the case of HD 115520
the distance as determined from Str\"omgren photometry is 140 pc
which leads an $M_V$ of 2.86 mag by using the calibrations of [8].
 On the other hand, the distance value of
130 pc estimated from a parallax of 3.29 $\pm$ 0.97 mas  provided by
the Hipparcos catalogue [9] yields  an $M_V$ of 1.02 mag which is
quite different from the photometric one. This ambiguity can be
explained by the uncertainties in the determination of each measured
distance. The large relative error ($\sigma (\pi))/ \pi \sim 0.30$ )
of the Hipparcos parallax for HD 115520 implies an $\sigma (M_V)  >
0.5$ mag, whereas in the present paper the uncertainty in the
apparent  magnitude  derived as explained in [2] from the standard
deviation of 579 data points of the two seasons gives an $m_V =
8.4305 \pm 0.0178$ (see Table 2) and an $\sigma (M_V)  < 0.1$ mag.
Although this latter value  does not include the uncertainty in
$M_V$ due to the photometric calibrations which can be as large as
0.3 mag for early type stars (e.g. [10]), we think that the
photometric distance is more reliable than the trigonometric one
because different photometric calibrations ([10] and [11]) lead to
similar distance values for HD 115520. Furthermore, similar values
of $m_V$ for HD 115520 have already been reported in previous papers
([12], [13], [1]). Therefore,  we will use the photometrically
determined distance to tray to establish the evolutionary status of
HD 115520.

\begin{figure}[t]
\begin{center}
\includegraphics[width=10cm]{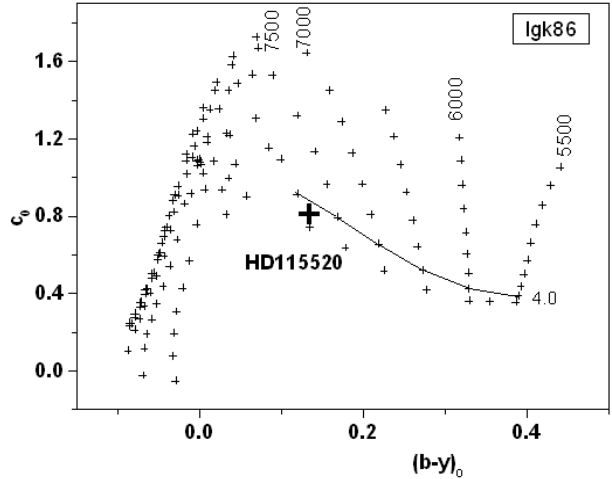}
\caption{Location of the photometric data of HD 115520 in the grids
of LGK86.}
\end{center}
\end{figure}

Figure 4 shows the observed position of HD 115520 (asterisk) in the
HR diagram and its associated uncertainty (cross upon the asterisk).
PMS and post-MS evolutionary tracks giving a range of masses between
1.45-1.60 $M_{\odot}$ for HD 115520 are shown with dotted and
continuous lines respectively. These evolutionary sequences were
computed by using the CESAM evolution code [14] with an input
physics appropriate to $\delta$ Scuti stars and a chemical initial
composition of $Z=0.013$ and $Y=0.28$. Also shown are the
theoretical pre-MS instability strip boundaries of the first three
radial modes obtained by [15].

According to the models depicted in Fig. 4  HD 115520 could either
be  in pre-MS stage with an age between 15-20 Myr or post-MS stag
with an age between 500-700 Myr. In the former case,  the age was
estimated as the time spent by the star travelling from the
birthline to the ZAMS in the HR diagram according to the isochrones
given by [16].

As shown by  [17] non-radial oscillation spectra in the low
frequency domain can be used to discriminate  between the pre- and
post-MS stage. In the present case, however, this is seldom possible
since the two detected peaks in HD 115520 are most likely due to
radial oscillations. In fact, we have tried to reproduce the
observed periods computing linear adiabatic pulsation models of HD
115520 for some selected pre- and post-MS models located within the
error box in Figure 4, but no satisfactory fit between observed and
theoretical frequencies was found. Therefore,  more observational
efforts are required to establish the true nature of this
interesting object.

\begin{table*}[!t]\centering
  \setlength{\tabcolsep}{1.0\tabcolsep}
 \caption{Reddening and unreddened parameters of HD 115520 }
  \begin{tabular}{lccccccccc}
\hline\hline
 $E(b-y)$ & $(b-y)_0$ & $m_0$ & $c_0$ & $V_0$ &  $M_v$ &  $DM $ & dst (pc) \\
\hline
0.000& 0.135 &0.170 & 0.807 & 8.43 & 2.68 & 5.75 & 141 \\
\hline
\end{tabular}
\end{table*}

\begin{figure}[t]
\begin{center}
\includegraphics[width=10cm]{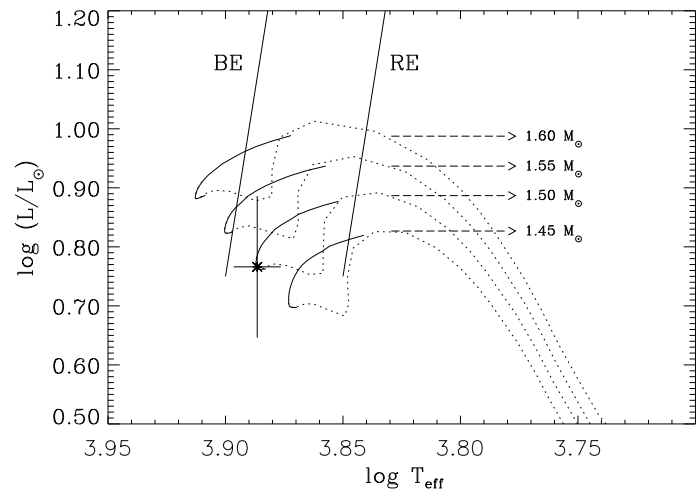}
\caption{Position of HD 115520 in the HR diagram.}
\end{center}
\end{figure}

\ack

This work was partially supported
by Papiit IN108106. JPS, RP,GM and BV thank the OAN for the facilities allowing the use of
the telescope time.

\end{document}